# Determination of Hydrated Phases of Nanosilica incorporated C3S and β-C2S using Nanoindentation Technique

Shaumik Ray[1,3], S. Gautham[2,3], Bala Pesala[1,3], Saptarshi Sasmal[2,3]

**Abstract:** Analysis of the mechanical properties of key phases of cementitious systems during hydration at the nanoscale is vital for a better understanding of the type and formation of key cement hydration phases. In this work, nanoindentation technique has been employed to calculate the volume fraction and elastic modulus of key hydrated phases, calcium silicate hydrate and calcium hydroxide during the hydration of nanosilica incorporated $C_3S$ and $β-C_2S$ at certain intervals for 50 days. The comparative analysis shows $β-C_2S$ has more dominant low-density calcium silicate hydrate, especially in the early stage of hydration. While in the case of $C_3S$, the dominant hydrated phase is high-density calcium silicate hydrate. Furthermore, the effect of nanosilica incorporation in the matrix of these cement constituents has also been demonstrated which shows a greater decrease of LD C-S-H and formation of more HD C-S-H for both C3S and $β-C_2S$. By tracking the mechanical properties of the other hydrated phase, calcium hydroxide, the pozzolanic activity of nanosilica has also been demonstrated. From the comparative analysis, it can be observed that the overall mechanical property of the constitutive phases is fairly similar.

Electronic mail:

[1]*Council of Scientific and Industrial Research (CSIR), Central Electronics Engineering Research Institute (CEERI), Chennai - 600113, Tamil Nadu, India*
[2]*Council of Scientific and Industrial Research (CSIR), Structural Engineering Research Centre (SERC), Chennai - 600113, Tamil Nadu, India*
[3]*Academy of Scientific and Innovative Research (AcSIR), Ghaziabad, India*

# 1.0 Introduction

Hydration of Ordinary Portland Cement (OPC) is a very complex phenomenon, due to the presence of multiple constituents in its matrix with varying crystal structures and impurities. This leads to a transitory nature of reactions, especially in the early stage of hydration. Hence understanding the characteristics of each of the cement constituents will be beneficial for a more detailed understanding of cement hydration. Of the cement constituents, tricalcium silicate ($C_3S$) and beta-dicalcium silicate ($\beta$-$C_2S$) are the two main constituents making up around 50 – 70% and 15 – 30% of the cement matrix respectively [1]. Hydration processes of the main cement constituents have been studied over the years, with a more focused analysis on the $C_3S$ due to its dominant presence in the cement matrix as well as its major contribution during the early stage hydration. Similarly, the accelerating effect of nanosilica incorporation in hydration of cement and CS has been extensively investigated using several techniques. In addition to the investigation of the structural and chemical changes during hydration of cementitious systems due to nanosilica incorporation using infrared (IR) spectroscopy, x-ray diffraction (XRD), nuclear magnetic resonance (NMR) as well as tracking the morphological variations using Scanning Electronic Microscopy (SEM) and Transmission Electron Microscopy (TEM), an investigation of the mechanical properties of cementitious systems is also vital at the micro- and nano-scales. These techniques give an overview of the structural, chemical and morphological changes during hydration but do not provide information regarding the mechanical properties. To address this, the experimental nanoindentation technique is very suitable as it works in three dimensions and also provides access to the mechanical properties of the indented material.

Nanoindentation technique evolved on the already established hardness testing method established by Brinell. The scale of the test was drastically reduced by using a needle to push into the surface of the material in question and interpreting the resistance provided by the material as its mechanical property [2], [3]. At the time of development and during the initial years, this experimental testing method was implemented on homogeneous material such as metal, glass, ceramic. This implementation was successful in determining the mechanical properties and also it was successful in exploring the microstructure making up the material [4]–[7]. The research on and with nanoindentation technique progressed only on homogeneous materials for a considerable duration of time as it was relatively easier to implement on such materials. In comparison, implementation and interpretation of nanoindentation on heterogeneous materials was relatively more complex due to the multiple material phases present. Generally, the heterogeneity of heterogeneous materials is non-uniformly spread throughout the material complicating the implementation of the nanoindentation technique as it is highly localized. Hence, there were more parameters such as the loading cycle, number of indents, interpretation of the tests to be considered while implementing complex heterogeneous materials. With continued research in this area, implementation of the nanoindentation technique on heterogeneous materials such as cementitious composites was also accomplished. Nanoindentation technique gained traction in the field of cementitious composites with its successful application to explore and analyze the microstructural composition and its mechanical properties of hydrated cement pastes [8]–[16].

This technique has immense potential, as it can be witnessed with its application by various researchers in the field of cementitious composites. As mentioned before, nanoindentation has been successful in analyzing the microstructural composition of the cement microstructure and has provided quantitative experimental proof to the presence of different hydration products such as multiple forms of calcium silicate hydrate (C-S-H) gels and other phases [13], [16], [17]. This technique has been successful in evaluating mechanical properties such as elastic modulus, hardness, creep, fracture toughness of the hydration products [18]–[20]. It has also been implemented to study the effect of blending admixtures such as fly ash, nanosilica, slag with OPC on its microstructural properties [8], [21]–[26]. With the relatively recent development in the application of experimental nanoindentation technique on cementitious composites, there are some areas about cement microstructure that are still unexplored. Since the research is relatively new, only simpler elastic properties such as elastic modulus, hardness has been explored. Studies have also been conducted on cement pastes blended with admixtures such as fly ash, nanosilica, etc. However, most of the studies focus on evaluating the properties once the cement paste has hardened with 28 days of hydration. Hence, there are areas such as early age hydration and complex mechanical properties still left to be explored. The study on the microstructure of cementitious systems during the early stages of hydration can help in better understanding of the hydration process and also the development of the mechanical properties. Having sufficient knowledge of the fundamentals of any process is always important to further any type of research and development. This is essential to better understand the macroscopic mechanical characteristics of the cementitious systems. This is done by carrying out nanoindentation analysis of these systems and the variation in mechanical properties due to incorporation of nanosilica in its matrix.

A significant amount of work has been carried out on the analysis of microstructural mechanical properties of cement due to varying water-cement ratio [27], [28]. Several other works which have investigated the microstructural mechanical properties of cement pastes with nanosilica, silica fume, fly ash, slag [29]–[35]. In these cementitious systems, by employing nanoindentation studies, significant mechanical properties have been obtained which can be used as a platform for better understanding and improvement of the macroscopic mechanical performances. Nanoindentation studies have been able to effectively differentiate between two distinct types of C-S-H in hardened cement pastes, the low-density C-S-H (LD C-S-H) and high-density C-S-H (HD C-S-H) [36]. This work has also mentioned that multiple studies using the statistical nanoindentation technique have demonstrated that the intrinsic elastic material properties of the two types of C-S-H are not dependent on the mix proportions of the cement-based materials. These two types of C-S-H are also known as the inner product and the outer product or low stiffness C-S-H and medium stiffness C-S-H. Furthermore, a comparative analysis of the mechanical properties of LD C-S-H and HD C-S-H have been shown in this work which shows significant variations in the reported mechanical properties of the two types of C-S-H in various literatures. A lot of work on the nanoindentation studies of nanosilica incorporated cement pastes have been carried out which shows the relative volume fraction of the HD C-S-H in total C-S-H increased due to the presence of nanosilica in the cement matrix. Further, a reduction in the average elastic properties of the hydration products were also observed due to nanosilica incorporation [37]. However, one of the key considerations is that for a better understanding of the effect of nanosilica in the mechanical properties of cement pastes, it is essential that an in-depth analysis on the microstructural mechanical properties of the individual cement constituents is carried out. The knowledge of the behaviour of every constituent phase separately will help in forming a reference basis to establish their effect and role in the hydrated cement material. The hydration process of the main calcium silicates of Portland cements, i.e., tricalcium silicate (C3S) and dicalcium silicate ($\beta$-$C_2S$), have been studied extensively over the years. However, more recently they have been gaining greater attention as there is a need to experimentally validate the models which are predicting the hydration of Portland cement and more importantly, the advances in characterization of C-S-H gels. Till date, there is no significant study that has carried out long term nanoindentation studies of the two main cement constituents, $C_3S$ and $\beta$-$C_2S$ and the effect of nanosilica incorporation. A study on the incorporation of nanosilica in the C3S matrix has been carried out for very early stage (till 24 hours) [38]. In this work, the particular mechanical properties of the hydrated phases, loose-packed (LP) C-S-H, LD C-S-H, HD C-S-H and CH have been extracted by statistical analysis/deconvolution technique. The analysis points towards the fact that the C-S-H formed from the hydration of cement and C3S are similar in nature. Furthermore, the cementitious phases during the early stage of hydration show greater proportion of LP C-S-H and LD C-S-H. With the progression in hydration, more formation of HD C-S-H was evident. In case of $C_3S$ with nanosilica incorporation, the nanoindentation analysis showed greater proportion of HD C-S-H and significantly lower proportion of LP C-S-H and LD C-S-H. Another comparative analysis of the mechanical properties of only $C_3S$ and $\beta$-$C_2S$ using nanoindentation studies have been carried out [39]. This study demonstrates that the C-S-H and CH formed in the hydrated $C_3S$ and $C_2S$ are same as those formed due to the hydration of cement. Further, in case of $C_3S$, the HD C-S-H and LD C-S-H were seen to be the dominant hydrated phases while for $C_2S$, LP C-S-H and LD C-S-H were more evident. As the hydration progressed, it was observed that the volume fraction of the various C-S-H phases shifted from LP C-S-H to LD C-S-H in case of $C_2S$ and from LD C-S-H to HD C-S-H in case of $C_3S$.

Therefore, this study is about the experimental investigation of the mechanical properties of $C_3S$ and β-$C_2S$ microstructures during their early stages of hydration to 50 days of hydration. Further, the effect of incorporating nanosilica in their matrix has also been explored. The experiment for this study has been the complex nanoindentation technique. It does pose some predicaments with the surface preparations, the number of tests to be conducted, but at the same time has its unique capabilities like assessing the mechanical properties at a highly localized region at the nanoscale. Hence, multiple indentations in the form of a pre-determined grid have been implemented to obtain multiple indentation data so that they can be statistically analyzed to obtain an acceptable understanding of the sample into consideration. These experiments and analysis have yielded some much-needed insight into the early age hydration of the major cement components.

## 2.0 Theoretical Background

### 2.1 Nanoindentation

The load vs. displacement graph of nanoindentation testing is obtained by the simultaneous measurement of the applied indenter load and the indenter displacement. The load is applied at a uniform rate with the indenter to a pre-determined maximum load. This maximum load is maintained for a certain duration (dwell time) and the indenter is unloaded at the uniform rate. These curves from this load cycle are utilized to determine the modulus of elasticity ($E$) and hardness ($H$) of the indented material. According to the Oliver and Pharr method (1992), the initial unloading contact stiffness is used to infer the elastic modulus of the material, $S = \mathrm{d}P/\mathrm{d}h$. This unloading contact stiffness is the slope of the initial portion of the unloading curve, as depicted in figure 1. In this figure, $h_{\max}$ is the depth beneath the specimen's free surface to a corresponding maximum load of $P_{max}$. The depth of residual impression is denoted by $h_r$, the depth of the contact circle by $h_c$ and the displacement due to the elastic recovery during unloading by $h_e$.

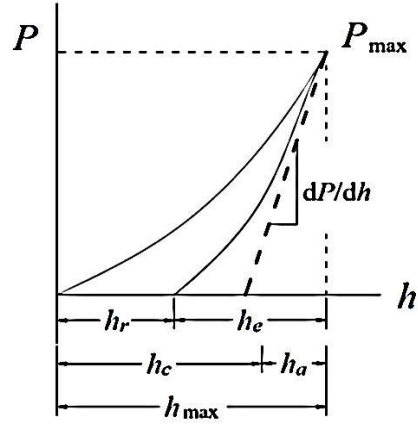

**Figure 1:** Plot of applied indenter load vs. indenter displacement

The slope of the unloading curve is considered as it represents the elastic properties of the specimen. By equating the projected area (*A*) that is in contact under the indenter to the area of punch, a solution is obtained as,

$$\frac{dP}{dh} = 2\sqrt{\frac{A}{\pi}} E_r \qquad (1)$$

where $dP/dh$ is the slope of the unloading curve; $E_r$ is the combined modulus which represents the elastic properties of both indenter and indented material. This equation 1 follows an analytical solution developed by Sneddon, who relates the elastic moduli of the indenter and material for penetration in elastic half-space as,

$$\frac{1}{E_r} = \frac{1-v^2}{E} + \frac{1-v_0^2}{E_0} \qquad (2)$$

where $E$ is the elastic modulus of the sample, $v$ is the Poisson's ratio of the sample, $E_0$ and $v_0$ are the elastic modulus and Poisson's ratio of the indenter. Berkovich indenter is the more generally used indenter and the projected area for it is given by $A = 3\sqrt{3}h_p^2 tan^2\theta$. Therefore, equation 1 can now be written as,

$$\frac{dh}{dP} = \frac{1}{2h_p}\sqrt{\frac{\pi}{24.5}}\left(\frac{1}{E_r}\right) \qquad (3)$$

where $dh/dP$, which is the reciprocal of the slope of the unloading curve, is known as compliance.

The applied load by the indenter, when divided by its projected area, is known to be equivalent to the hardness of the indented sample.

$$H = \frac{P}{A} \tag{4}$$

**2.2 Gedanken guidelines for performing experimental grid nanoindentation**

It is understood that single nanoindentation will result in an evaluation of the mechanical property of the indented material at that location. Hence to obtain a more comprehensive and realistic picture of the heterogeneous material in question, multiple indentations in the form of a well thought out grid is highly recommended. The guidelines that need to be followed for executing multiple nanoindentations on heterogeneous materials are elucidated in the Gedanken experiment. According to the findings stated by the experiment, if the indentation depth is small, roughly *h/D < 1/10* (*h* is the indentation depth and *D* is the length scale of the indented phase), then the resulting data will be that of a particular phase. With the small indentation depths, the potential volume fractions can be estimated as,

$$f_J = \frac{N_J}{N}; \sum_{J=1}^{n} N_J = N \tag{5}$$

where *n* is the number of indentations occurring on a particular material phase *J* is given by $N_J$. The different material phases can be identified by the varying values of material properties. Hence, $f_J$ is the volume fraction of mechanically identifiable individual phase of the material. Now, if homogenized material properties of two or more phases are to be evaluated with the indentation, then larger indentation depths, in general, *h/D > 6* was found to be suitable. When multiple indentations are to be conducted, sufficient distance should be maintained between the individual indents. And the suitable minimum spacing (*l*) suggested by the experiment is $l \gg D/\sqrt{N}$. This distance of spacing was found to be sufficient enough to avoid any interference creeping in from the neighbouring indents (Figure 2).

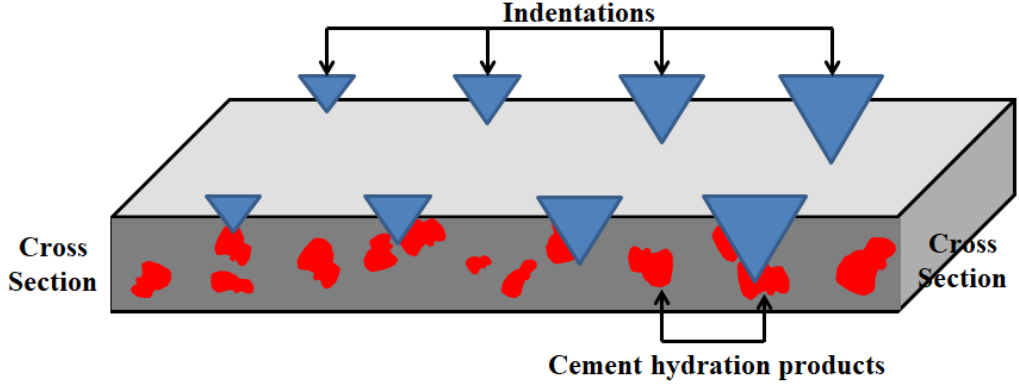

**Figure 2:** Representative diagram for grid nanoindentation and the effect of depth of indentation on measured mechanical properties

## 2.3 Deconvolution technique for analyzing the data obtained from experimental grid nanoindentation

Once the guidelines to perform grid nanoindentation established, there is a need for statistically analyzing the large array of data that is obtained. The properties of the individual mechanical phases can be obtained by performing statistical deconvolution on the obtained data. This analysis divides the available data into different groups and presents with the mean and standard deviation of the values in each groups. This analysis can be implemented on any mechanical property such as elastic modulus, hardness, etc.

Deconvolution starts with building histograms using equally spaced $N$ bins of size $b$ from all the data obtained from grid nanoindentation. Each of these $N$ bins should be assigned with a frequency of occurrence $f_i$. This frequency of occurrence has to be normalized with the total number of obtained data as $f_i/N$. Thereafter, the probability density function (PDF) can be calculated as a set of discrete values,

$$P_i = \frac{f_i}{Nb} \tag{6}$$

Finding $j = 1…M$ individual PDFs related to a single material phase is what deconvolution does. Now, the PDF for a single material phase, assuming Normal distribution (Gauss distribution), can be written as,

$$p_j(x) = \frac{1}{\sqrt{2\pi s_j^2}} exp \frac{-(x - \mu_j)^2}{2s_j^2} \tag{7}$$

where $\mu_j$ is the mean values of the mechanical property of the $j$-th phase and $s_j$ is its standard deviation. These mean and standard deviation are computed from $n_j$ number of values as,

$$\mu_j = \frac{1}{n_j}\sum_{k=1}^{n_j} x_k \qquad s_j^2 = \frac{1}{n_j-1}\sum_{k=1}^{n_j}(x_k - \mu_j)^2 \tag{8}$$

where $x$ is the approximated mechanical property ($E$, $H$, etc.). Now, the PDF that will cover all the $M$ phases can be written as,

$$C(x) = \sum_{j=1}^{M} f_j p_j(x) \tag{9}$$

where $f_j$ is the volume fraction of a single material phase given by, $f_j = n_j/N$. The individual distributions are then evaluated by minimizing the error function given as,

$$\min \sum_{i=1}^{N^{bins}} [(P_i - C(x_i))P_i]^2 \tag{10}$$

This minimization of the error function is governed by the random Monte Carlo generation of total $M$ PDFs which have to satisfy the following compatibility condition,

$$\sum_{j=1}^{M} f_j = 1 \tag{11}$$

## 3.0 Materials and methods used in the experimental nanoindentation studies

### 3.1 Materials

Synthetic C$_3$S and β-C$_2$S purchased from M/s. Sarl Mineral Research Processing, France has been used in this study. The nanosilica has been purchased from Nanostructured and Amorphous Materials, Inc., USA, and has an average particle size of 80 nm with a purity of 99%.

### 3.2 Sample preparation

The mix was cast into small moulds of size after mixing the C$_3$S and β-C$_2$S with deionized water at a water-binder ratio of 0.6. For the nanosilica samples, 5% nanosilica (by weight) was mixed with the cementitious systems. At pre-determined stages of hydration, that is, 3, 7, 14, 28 and 50 days after mixing, the samples were demoulded. For the β-C$_2$S sample, day 3 was not considered because at such an early stage of hydration, the hardness of β-C$_2$S samples was found to be significantly low. Thus, the samples were seen to degrade during the coarse and fine polishing stage of sample preparation which is crucial to carry out the nanoindentation experiments. Since one of the cruxes of experimental nanoindentation is the complex time-consuming sample preparation, the demoulded samples were immersed completely in acetone for ten days to arrest the hydration by removing the pore water [40]. Since the sample sizes are very small, after arresting the hydration, the samples were impregnated with an epoxy resin to facilitate the sample preparation procedure. After the impregnation, a thin layer was cut off from the top to reveal the sample surface using a Buehler Isomet 5000 precision cutting diamond saw. This was followed by cleaning of the samples in a bath ultrasonicator filled with acetone. The freshly cut surface is then polished to obtain a smooth surface.

Obtaining a smooth surface on a heterogeneous brittle material requires considerable attention to detail and is not an easy process. Several researchers have implemented different variations of certain procedures to obtain a sufficiently smooth surface [16], [20], [41], [42]. In all the cases, the roughness of the surface obtained was measured by using an in-situ imaging method and the measured surface roughness was expressed in the form of root mean square (RMS) roughness number. By observing the different surface preparation procedures implemented by various researchers it was concluded that the surface preparation needed two main steps, i.e., coarse polishing and fine polishing. Coarse polishing involved grinding with silicon carbide abrasive papers of different fineness; fine polishing involved polishing with diamond suspensions of different fineness. From different trial and error attempts the following surface preparation procedure was proposed and utilized, Table 1. The samples were cleaned in a bath ultrasonicator in between and after the steps mentioned in Table 1.

**Table 1:** Details of the procedure followed to prepare the sample surface for nanoindentation

| Sample | | Preparation Method | |
|---|---|---|---|
| Feature | Size | Coarse polishing (Silicon carbide abrasive paper) | Fine polishing (Diamond suspension) |

| C3S and β-C2S (control sample) and 5% nanosilica incorporated w/c ratio = 0.6 50 days | Diameter = 30 mm  Height = 6 mm | 600 grit – 3 hours  1200 grit – 3 hours | 9 μ - 1 hour  6 μ - 1 hour  3 μ - 2 hours  1 μ - 1.5 hours |
|---|---|---|---|

### 3.3 Grid nanoindentation on cementitious composites

Once the sample surfaces are satisfactorily prepared, the subsequent step is to design and execute multiple indentations in a grid layout. Nanoindentation testing yields a highly localized and homogeneous property of the material present at the location of testing. Hence, to implement nanoindentation testing on a highly heterogeneous material such as cementitious systems, it necessary to test multiple locations to obtain a realistic picture [14], [43], [44]. This execution demands careful planning of the tests. For this, the Gedanken experiment is considered and the type of material is considered for pre-determination of the parameters such as grid size to cover the inherent phases, grid spacing to distinguish the adjacent test results, and the maximum loading to ensure that the individual phase properties are represented.

However, these careful planning of the parameters will not guarantee the achievement of proper and desired results. The complex heterogeneity of the cement microstructure will pose some problems to the nanoindentation testing. This intricate microstructure results in the indeterminacy of phase present under the load-bearing indenter which may result in the indentation of unfavourable phases such as pores, defects and/or impurities. As a result, there may be some irregular load vs. indentation plots which need to be sorted out before analysis. A flat region at the beginning of the curve, flat shoulders in between the curves are some such instances which have to be looked out for.

### 3.4 Experimental nanoindentation setup

With the aforementioned parameters carefully considered and by reviewing established works [14], [43], [44], the parameters were carefully chosen as a grid of size 10 × 10 points was adopted with 5 µm as the grid spacing between the indents; the load was applied at a rate of 1200 mN/min till a maximum load of 2000 mN; dwell time for 10 seconds after which the unloading was done at the same rate as loading, figure 3. With these parameters, experimental grid nanoindentation was performed on the samples at different stages of hydration. The nanoindenter available at CSIR-SERC, manufactured by CSM Instruments, Switzerland was utilized to conduct the tests.

## 4.0 Results

Conducting the nanoindentation tests yielded the corresponding load vs. displacement curves which were used to analyze the mechanical properties of the pastes. The load case implemented and the typical load vs. displacement curve obtained from the experimental nanoindentation is as shown in Figure 33. The mechanical properties are evaluated from the initial linear portion of the unloading the curve. The elastic modulus values are calculated as explained in the aforementioned section 2.1.

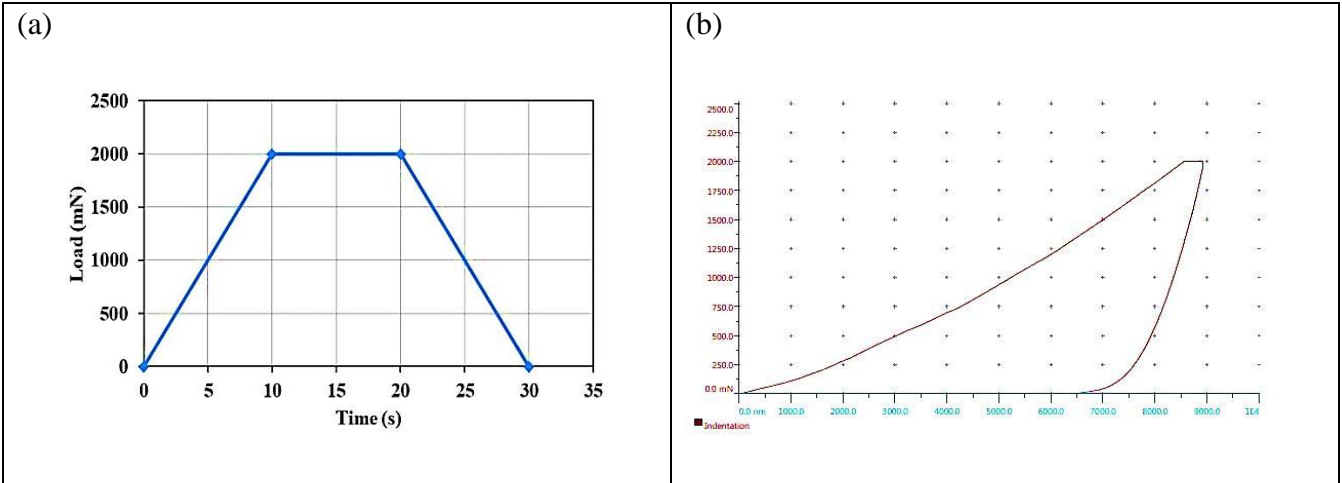

**Figure 3:** (a) Graphical representation of the loading and unloading steps for indentation (b) A typical load vs. displacement curve obtained from experimental nanoindentation

Care has to be taken to eliminate faulty indentations by observing the curves for irregularities such as the ones mentioned in section 3.3. The phases of interest in this study are LD C-S-H, HD C-S-H and CH. Hence by observing the indentation depth values reported in the literature [16], [17], [20], curves exhibiting indentation depth more than 500 nm were eliminated.

The load vs. displacement curves remaining after the filtration was used to calculate the elastic modulus and hardness values. The mechanical properties obtained from grid nanoindentation were statistically analyzed using the deconvolution technique into different phases. This statistical analysis provided the elastic modulus of each phase and the area under each phase representing the corresponding volume fraction. This analysis was done for each sample including the samples of different composition along their different stages of hydration. The deconvolution curves obtained for the nanoindentation results of $C_3S$ and $\beta$-$C_2S$ along its stages of hydration are as shown in figure 4 and Figure 5. This validation of the evaluated values with the reported values helps in bolstering the experimental procedure implemented and thereby the results of the rest of the tests are presented with confidence.

**C$_3$S and 5% nanosilica incorporated C$_3$S**

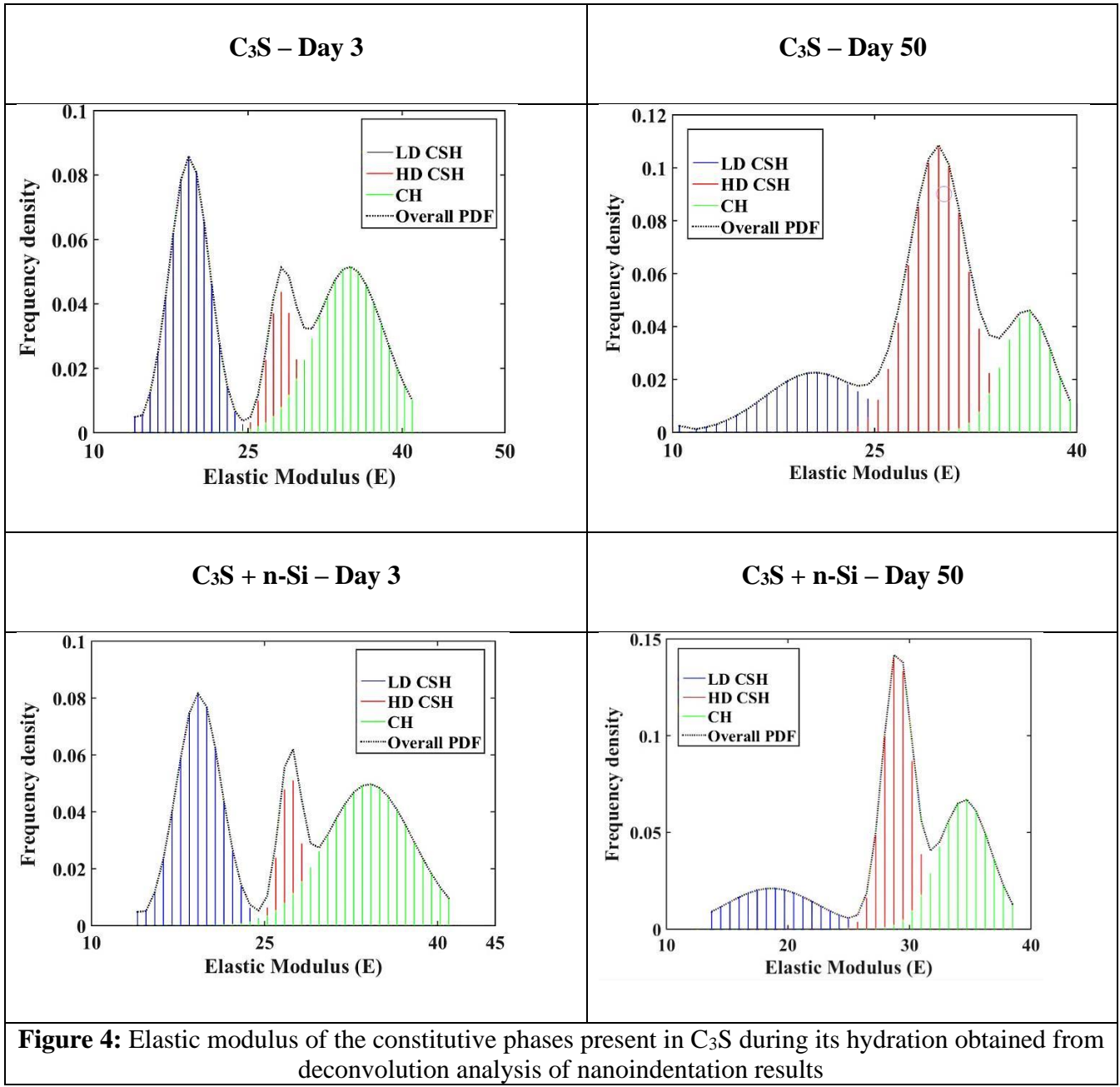

**Figure 4:** Elastic modulus of the constitutive phases present in C$_3$S during its hydration obtained from deconvolution analysis of nanoindentation results

**β-C$_2$S and 5% nanosilica incorporated β-C$_2$S**

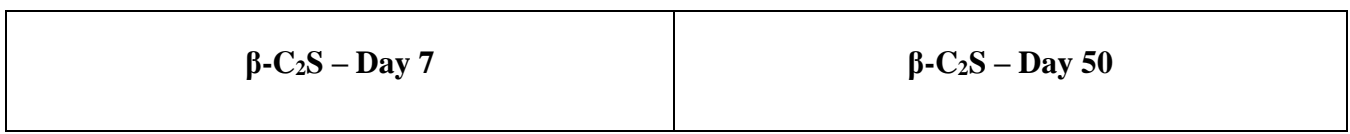

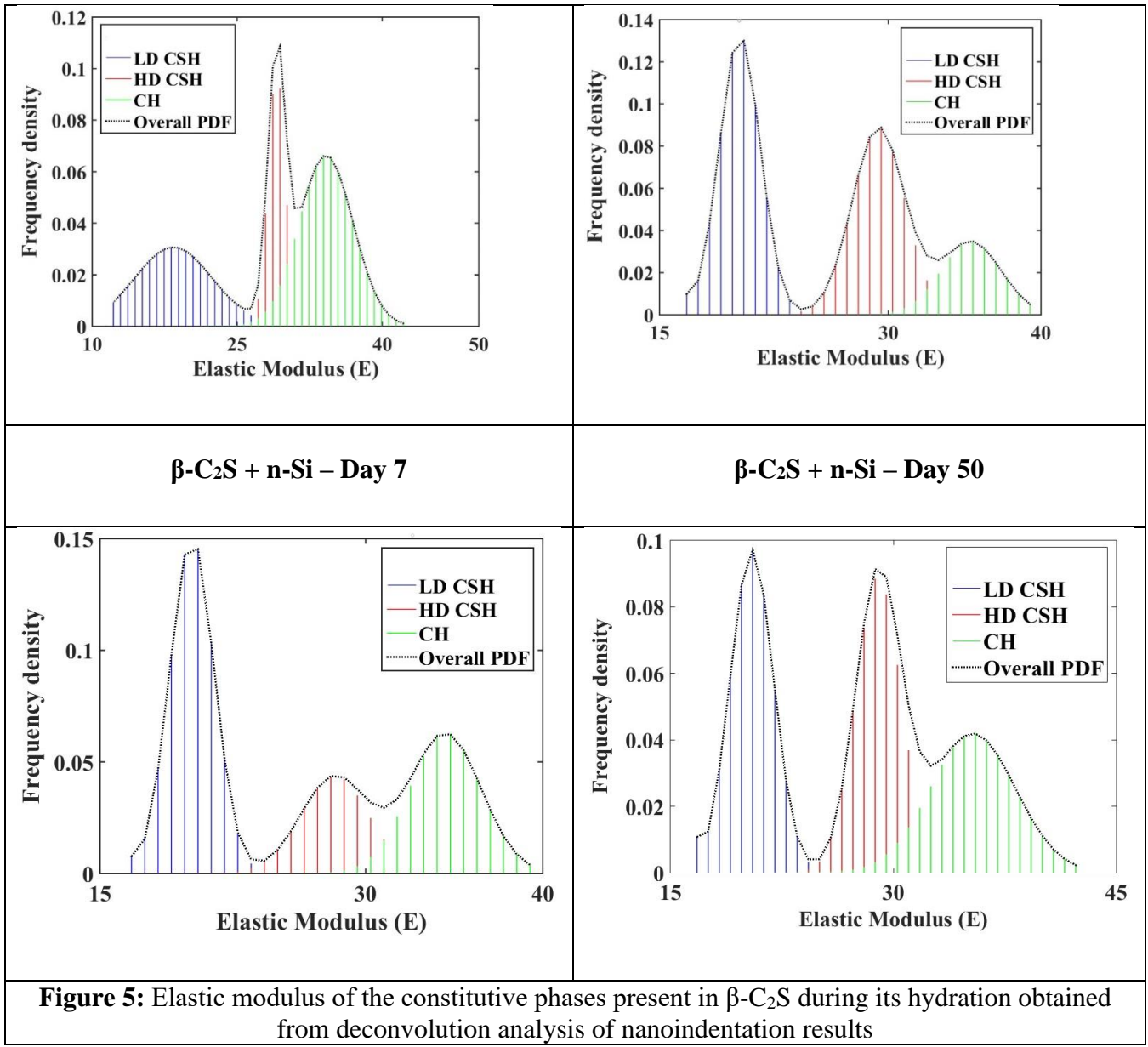

**Figure 5:** Elastic modulus of the constitutive phases present in β-$C_2S$ during its hydration obtained from deconvolution analysis of nanoindentation results

The elastic modulus and the corresponding volume fractions present the four different mixes ($C_3S$, $C_3S$ + n-Si, β-$C_2S$, β-$C_2S$ + n-Si) along its stages of hydration are presented in Table 2.

**Table 2:** Summary of the statistical nanoindentation test results

| Property | Specimen | $C_3S$ | | | | | $C_3S$ + nanosilica | | | | |
|---|---|---|---|---|---|---|---|---|---|---|---|
| | Age | 3 day | 7 day | 14 day | 28 day | 50 day | 3 day | 7 day | 14 day | 28 day | 50 day |
| LD CSH | Modulus, GPa | 19.3±6.6 | 20.6±5.1 | 18.6±6.3 | 17.8±8.3 | 17.8±8.3 | 18.2±6.8 | 17.5±6.7 | 17.9±7.5 | 19.3±7.3 | 19.8±7.1 |
| | Volume, | 51% | 45% | 40% | 38% | 35% | 50% | 41% | 37% | 34% | 33% |

| | % | | | | | | | | | |
|---|---|---|---|---|---|---|---|---|---|---|
| HD CSH | Modulus, GPa | 28.5±5.4 | 27.2±6.5 | 29.7±7.8 | 26.9±7.5 | 26.9±7.5 | 29.3±6.2 | 26.9±7.3 | 27.5±6.9 | 27.6±7.7 | 29.3±6.5 |
| | Volume, % | 13% | 23% | 36% | 42% | 46% | 9% | 31% | 42% | 49% | 53% |
| CH | Modulus, GPa | 38.1±6.2 | 40.6±7.8 | 37.9±5.6 | 36.6±7.2 | 36.6±7.2 | 40.1±8.4 | 37.4±6.8 | 38.2±5.8 | 38.8±6.5 | 37.4±6.3 |
| | Volume, % | 36% | 32% | 24% | 20% | 19% | 41% | 28% | 21% | 17% | 14% |

| Property | Specimen | β-$C_2S$ | | | | β-$C_2S$ + nanosilica | | | |
|---|---|---|---|---|---|---|---|---|---|
| | Age | 7 day | 14 day | 28 day | 50 day | 7 day | 14 day | 28 day | 50 day |
| LD CSH | Modulus, GPa | 15.6±8.7 | 16.2±7.6 | 17.5±8.3 | 17.6±7.4 | 16.2±8.1 | 16.8±7.3 | 15.7±7.7 | 17.3±7.5 |
| | Volume, % | 71% | 67% | 59% | 48% | 67% | 59% | 48% | 42% |
| HD CSH | Modulus, GPa | 25.7±7.5 | 25.9±7.2 | 26.4±6.8 | 26.1±6.7 | 25.1±6.9 | 26.3±7.6 | 27.1±7.4 | 26.6±7.0 |
| | Volume, % | 18% | 21% | 26% | 35% | 14% | 17% | 30% | 39% |
| CH | Modulus, GPa | 36.2±7.1 | 35.8±6.9 | 35.5±7.5 | 36.1±7.7 | 35.7±7.5 | 27.1±7.4 | 36.9±6.8 | 37.4±8.2 |
| | Volume, % | 11% | 12% | 15% | 17% | 19% | 30% | 22% | 19% |

From the evaluated results, the volume fractions of the constitutive phases present in the four mixes are observed and plotted as in figure 6.

From figure 6, it can be observed that the volume fraction of LD C-S-H gradually decreases as the hydration progresses and correspondingly the volume fraction of HD C-S-H increases. This is because HD C-S-H is more densely packed LD C-S-H; as more and more LD C-S-H is being produced as the result of hydration, it collapses on itself due to its weight and due to lack of space and gets densely packed as HD C-S-H. The volume fraction of CH also keeps reducing as the hydration progress and this can be explained by the fact that some of the CH get used in the secondary hydration reaction to form LD C-S-H. Also, as evident from the comparative analysis, the volume fraction of LD C-S-H for β-$C_2S$ was seen to be more for the LD C-S-H of $C_3S$. Simultaneously, it is also evident that the formation of HD C-S-H in C₃S is faster as compared to β-C₂S. For the nanosilica incorporated samples, it was seen that the trend of reduction of LD C-S-H and formation of HD C-S-H is more for both the cement constituents. This can conclusively demonstrate the fact that the rates of formation and the type of hydration products formed for $C_3S$ and $\beta-C_2S$ is different and has varying mechanical properties. The effect of nanosilica is also clearly evident from the comparative analysis. The reduction in the CH formation is more evident in the case of $C_3S$ and the role of nanosilica as a pozzolanic material is seen. However, this prominent pozzolanic activity was not evident in the case of $\beta-C_2S$, the reason for which needs further studies and investigation.

This is observed for both $C_3S$ and $\beta-C_2S$ concerning the trend of volume fractions of constituent phases as hydration progresses. The addition of nanosilica into the mix shows a decrease in the volume fractions of LD C-S-H and CH, and correspondingly an increase in the volume fraction of HD C-S-H in both $C_3S$ and $\beta-C_2S$ pastes. Another prominent observation in the comparative analysis was that the presence of nanosilica resulted in a greater reduction of LD C-S-H with more formation of HD C-S-H. However, across the different mixes, it can be observed that the overall mechanical property (elastic modulus) of the constitutive phases is fairly similar. This further strengthens the fact that the elastic modulus is an intrinsic property and is not affected by the stage of hydration or incorporation of nanosilica.

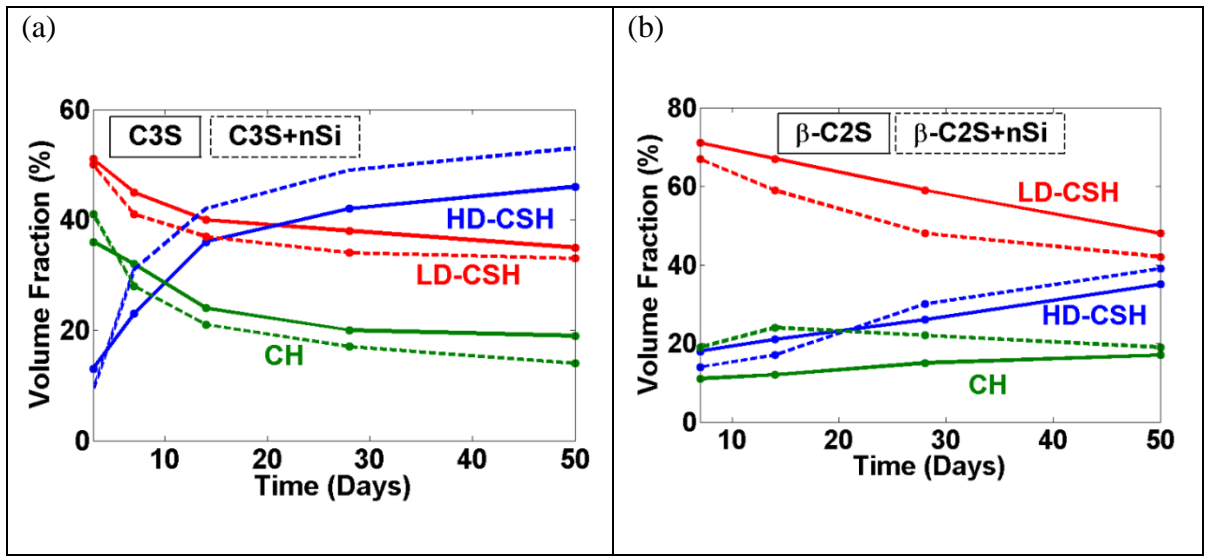

**Figure 6:** Volume fractions of the constitutive phases present in the four mixes along their stages of hydration for (a) $C_3S$ and (b) $\beta-C_2S$

**5.0 Conclusions**

From the experimental study conducted, it is proven that nanoindentation can be effectively implemented on the major cement constituents, $C_3S$ and $β-C_2S$, across their different stages of hydration as well as the effect of nanosilica can be effectively tracked. This has facilitated in exploring and observing their microstructure as the hydration progresses thereby providing a unique insight into the early-stage development of the major components present in cement. The results shown here clearly demonstrate the varying mechanical properties of $C_3S$ and $β-C_2S$ during the first week of hydration to 50 days. From the comparative analysis the key following observations were demonstrated:

- At an early stage of hydration (around 7 days), the presence of LD C-S-H was more evident in case of $β-C_2S$ as compared to $C_3S$ which was gradually seen to decrease for both the cement constituents.
- Presence of HD C-S-H was more in case of $C_3S$ as compared to $β-C_2S$, especially at later stages of hydration.
- The presence of nanosilica in the matrix of $C_3S$ and $β-C_2S$ resulted in a greater decrease of LD C-S-H and formation of more HD C-S-H.
- The reduction in CH, especially due to the presence of nanosilica was evident confirming pozzolanic reactions during hydration, which was more prominent in the case of $C_3S$. Some ambiguity was observed in case of $β-C_2S$ which needs investigation in future studies.

**Acknowledgements** Shaumik Ray would like to thank CSIR-Senior Research Fellowship for financial support.